\documentclass[twocolumn,secnumarabic,amssymb, amsmath, nofootinbib,tightenlines, nobibnotes, color, aps, prl,epsfig]{revtex4}
\usepackage{graphicx}
\usepackage{dcolumn}
\usepackage{bm}
\begin{document}
\preprint{APS/123-QED}
\title{ Top cross section in the LHeC and FCC-he energy range }

\author{G.R.Boroun}%
 \email{boroun@razi.ac.ir }
\affiliation{ Department of Physics, Razi University, Kermanshah
67149, Iran}

\date{\today}
\begin{abstract}
Predictions of the top-pair production cross section in
electron-proton (ep) collisions at the future circular collider
hadron-electron (FCC-he) and the large hadron electron collider
(LHeC) in the leading order (LO) and the next-to-leading order
(NLO) approximations in perturbative quantum chromodynamics (QCD)
are presented. Numerical results for the top structure functions
and the ratio of structure functions are computed at the
renormalization scale $\mu^{2}=Q^{2}+4m^{2}_{t}$ in the collinear
generalized double asymptotic scaling (DAS) approach. These
results are presented in terms of the effective parameters of the
parameterization of $F_{2}(x,Q^{2})$, where relying on a
Froissart-bounded. These quantitative results can be important for
future collider experiments at the center-of-mass energy frontier
and the improvement of the phenomenological models for the
development of the cosmic ray cascades in ultra-high-energy
domain. We obtained the uncertainties because of the power-law
behavior $x^{-\Delta}$ of the collinear parton distribution
functions (PDFs), the factorization, and renormalization scales,
on the top properties at high inelasticity
for the FCC-he center-of-mass (CM) energy at the NLO approximation.\\
\end{abstract}
 \pacs{***}
\maketitle
\subsection{I. INTRODUCTION}

Top physics at future electron-proton colliders will be
highlighted to show the future progress in the large hadron
electron collider (LHeC) and future circular collider
hadron-electron (FCC-he) accelerators [1,2]. The main goal of the
future at the LHeC and FCC-he is to achieve a deeper knowledge of
the hadronic structure at high energies through the deep inelastic
scattering (DIS) process, where an electron emits a virtual photon
that interacts with a proton target. In particular, the proton
structure can be studied through the $\gamma^{*}p$ interaction,
with the behavior of the observables being determined by the QCD
dynamics at high energies. These ep colliders (i.e., LHeC and
FCC-he), with those smaller backgrounds, can be of great
importance in illuminating the nature of the new physics for
top-pair production. The top quark can be produced by the LHeC and
FCC-he colliders in $\gamma^{*}p{\rightarrow}t\overline{t}X$
reactions [3]. For the first time, at the LHeC, the top-pair
production can be studied in DIS. According to the standard model
(SM), top -pair is produced dominantly in gluon-boson fusion at
$x<0.1$. The LHeC opens top quark parton distribution function
(PDF) physics as a new field of research. Top -quark production
could not be observed at hadron-electron ring accelerator (HERA),
but it will become a major topic of precision physics and
discovery
in the LHeC and FCC-he colliders. \\
A single top is produced in charged currents (CC), with a total
cross section of order 5 pb, which can easily be estimated from
the leading -order (LO) calculation of Wb scattering [2] or 1.89
pb as reported in Ref.[1] at a center-of-mass energy of 1.3 TeV,
i.e., with an electron-beam energy of 60 GeV and an LHC proton
beam of 7 TeV. In CC, this leads to a top-beauty final state,
while in neutral currents (NC) this gives rise to a pair producing
top-antitop quarks in photoproduction and DIS. Photoproduction of
the top-antitop quark pairs
($\gamma{g}{\rightarrow}t\overline{t}$) is the second most
important production mode for top quarks at the LHeC as a total
cross -section of 0.05 pb is expected at the LHeC. Furthermore,
the DIS regime of $t\overline{t}$ production will allow us to
probe the top cross section. In DIS, the total cross section for
top quark pair production mode is $\sim 53$ fb at the LHeC and
$\sim 1$ pb at the FCC-he [4]. Here, it may be useful to refer to
older discussions about heavy-quark production. The authors in
Refs.[5] and [6] studied the top quark production in ep-collisions
at HERA energies. In electron -proton colliders like HERA, a heavy
quark pair is created via the boson-gluon fusion (BGF) mechanism
[5], where the electromagnetic coupling in
$ep{\rightarrow}t\overline{t}X$ pair production is defined by the
following interaction lagrangian density
\begin{eqnarray}
\mathcal{L}_{em}=-\sqrt{4\pi
\alpha_{EM}}\bigg{[}e_{e}\overline{\Psi}_{e}{\gamma
^{\mu}}\Psi_{e}+e_{t}\overline{\Psi}_{t}{\gamma
^{\mu}}\Psi_{t}\bigg{]}A_{\mu}.
\end{eqnarray}
Here, the Dirac fields $\Psi_{e}$, $\Psi_{t}$ and the vector field
$A_{\mu}$  describe the electron, the top and the photon,
respectively. In Eq.1, $e_{e}$ and $e_{t}$ give, respectively, the
electric charge of electron and top -quark in units of $\sqrt{4\pi
\alpha_{EM}}$, where $\alpha_{EM}$ is the fine structure constant.
The cross section in the BGF is given by the formula
\begin{eqnarray}
d\sigma=\frac{1}{8}\frac{1}{4\mathbf{p}.\mathbf{l}}G(z,s)dzdPS^{(3)}\sum_{\mathrm{spins}}|\mathrm{M}_{t}|^2.
\end{eqnarray}
Here, $G(z,s)$ is the gluon distribution where the momentum scale
of the gluon density has been chosen to correspond to the total
invariant energy $s$ of the produced top quark system.
$\mathbf{p}$ and $\mathbf{l}$ are momentum of gluon and lepton in
the subprocess, and $dPS^{(3)}$ is the three-particle phase space.
The exact solution of the total cross section is described in
Refs.[5] and [6]. A useful approximate form of the exact total
cross -section for the large values of $\sqrt{s}$ is provided by
the Weizs$\ddot{a}$cker-Williams (WW) approximation [7], which
provides a convenient framework to derive simple approximate forms
of Eq.(2), as
\begin{eqnarray}
\sigma=\int_{y_{\mathrm{min}}}^{1}dy
P_{\gamma}(y)\int_{z_{\mathrm{min}}}^{1}dz
G(z,M_{p}^{2})\widehat{\sigma},
\end{eqnarray}
where $P_{\gamma}$ is the photon -splitting function and
$\widehat{\sigma}$ is the cross section of the $t\overline{t}$
pair production by a real photon and
$z_{\mathrm{min}}=(4m_{t}^{2}+Q^{2})/(ys)$. The WW approximation
agrees with the exact pair production for top quarks as discussed
in the literature [5,6]. The corresponding hadron-level cross
sections, $\sigma^{t}_{k=2,L}$ in the fixed-flavor-number scheme
(FFNS), haves the form [8]
\begin{eqnarray}
\sigma^{t}_{k}=\int_{z_{\mathrm{min}}}^{1}dz
G(z,\mu_{F})\widehat{\sigma}^{t}_{k,g}(\frac{x}{z},\mu_{F},\mu_{R}),
\end{eqnarray}
where $G=xf_{g}$ is the gluon distribution function of the proton.
$\mu_{R}$ and $\mu_{F}$ are the renormalization and factorization
scales, respectively. Here, $\widehat{\sigma}^{t}_{k,g},~(k=2,L)$
are the $\gamma^{*}g$ cross sections as at LO approximation, the
photon -gluon  component of the heavy-quark leptoproduction is
described by the following parton-level interaction:
\begin{eqnarray}
\gamma^{*}+g{\rightarrow}t+\overline{t}.\nonumber
\end{eqnarray}
The leptoproduction cross sections $\sigma^{t}_{k=2,L}(x,Q^2)$ are
related to the structure functions $F^{t}_{k}(x,Q^{2})$ as
follows:
\begin{eqnarray}
F^{t}_{L}(x,Q^{2})&=&\frac{Q^2}{8\pi^2\alpha_{EM}x}\sigma^{t}_{L}(x,Q^2),\nonumber\\
F^{t}_{2}(x,Q^{2})&=&\frac{Q^2}{4\pi^2\alpha_{EM}}\sigma^{t}_{2}(x,Q^2).
\end{eqnarray}
In the case of unpolarized initial states and neglecting the
contribution of Z-boson exchange, the structure functions are
related to the double differential cross section by the following
form:
\begin{eqnarray}
\frac{d^{2}\sigma^{{t}\overline{{t}}}}{dxdQ^{2}}&=&\frac{2\pi
\alpha_{EM}^{2}}{xQ^{4}}\bigg{\{}[(1+(1-y)^{2}]F_{2}^{t}(x,Q^{2})\nonumber\\
&&-y^{2}F_{L}^{t}(x,Q^{2})\bigg{\}},
\end{eqnarray}
where $y=Q^{2}/sx$ is the inelasticity with s the ep center-of-mass energy squared.\\
Fermion pair production in the DIS is sensitive to the gluon
density in the proton. By using DIS $t\overline{t}$ production,
one can study the top component of the structure function at the
small $x$ region [9-10]. In this domain, the power-law behavior of
the gluon distribution function
$xf_{g}(x,Q^{2}){\propto}x^{1-\Delta}$ measures the intercept of
the BFKL trajectory $\Delta=\Delta_{\mathrm{BFKL}}(t=0)$. The
standard parameterization of the gluon distribution function for
$x{\rightarrow}0$ is given by
$$
xf_{g}(x,Q^{2})|_{x{\rightarrow}0}=A_{g}(Q^{2})x^{-\Delta},
$$
where $A_{g}$ is $Q^{2}$ dependent and $\Delta$ is strictly
positive. The gluon exponent at low values of $x$ at
$Q^{2}=1~\mathrm{GeV}^{2}$ has been obtained by MSTW08 NLO is
$0.428^{+0.066}_{-0.057}$ [11]. The effective exponent for the
gluon distribution at $Q^{2}=10 ~\mathrm{GeV}^{2}$ and $x=10^{-4}$
obtained by NNPDF3.0, CT14, MMHT14, ABM12, and CJ15 had values of
0.20, 0.15, 0.29, 0.15, and 0.14, respectively. The value obtained
by fixed coupling LLx BFKL gives $\Delta{\simeq}0.5$, which is the
so-called hard-Pomeron exponent. The gluon exponent is determined
and applied to the deep inelastic lepton nucleon scattering at low
values of $x$ in Ref.[12]. Ref.[13] used a form inspired by the
double asymptotic $x,Q^{2}$ scaling. The hard -Pomeron behavior of
the photon -proton cross section based on a simple power-law
behavior and double asymptotic scaling (DAS) at low $x$ values for
$10<W<10^{4}~\mathrm{GeV}$ (where $W^{2}$ is the invariant mass
squared of the hadronic system in the final state) in the
kinematic range of very high energy electron -proton/ion
collider(VHEep) is shown in Ref.[14]. The authors in Ref.[15]
presented a tensor-Pomeron model, where it was applied to low-$x$
deep inelastic lepton-nucleon scattering and photoproduction
processes. In this model, in addition to the soft tensor Pomeron,
a hard tensor Pomeron and Reggeon exchange was included. In this
case, the hard-Pomeron intercept was determined to be
$0.3008(^{+73}_{-84})$ with the latest HERA data for $x<0.01$.\\
In the very small $x$ domain, one studies QCD at a high gluon
density where ep collider kinematics reaches a maximum of
$Q^{2}{\simeq}1~\mathrm{TeV}^{2}$ and $x{\simeq}10^{-5..-6}$ for
LHeC and $10^{-7}$ for FCC-he. The ep center-of-mass (CM) energy
at the LHeC and FCC-he ranges reaches up to $\sqrt{s}\simeq 1.3$
and
$3.5~\mathrm{TeV}$, about 4 and 10 times the CM energy range of ep
collisions at HERA, respectively.\\
In this paper, we consider the top quark pair production by the
collinear generalized double asymptotic scaling approach [16] at
the LHeC and FCC-he in the 5-flavor scheme (5FS) at the LO and NLO
approximations in QCD [17]. The interest in a measurement of the
top pair production, especially at low $x$, is related to the
determination of the gluon distribution. The gluon distribution
function is directly related to the proton structure function.
Theoretical analysis of the parameterization of the proton
structure function at low $x$, in the context of the fulfillment
of the Froissart prescriptions, is suggested in Ref.[18]. The
authors in Refs. [19,20] derived the longitudinal structure
function based on this method. We focus on the parameterization of
the top structure function for $Q^{2}<m_{t}^{2}$ and study the
collinear approach to the parameterization of the top structure
function in a typical region of the LHeC and FCC-he extended from
the HERA kinematics. In Ref.[18], the authors fitted the HERA DIS
data on $F_{2}^{p}$ at low $x$ in a wide range of the momentum
transfer ($1~\mathrm{GeV}^{2}<Q^{2}<10^{5}~\mathrm{GeV}^{2}$). We
use these kinematics according to the ep collider CM energies at
high
inelasticity.\\
Figure 1 shows the area used for the transition from the HERA to
the LHeC and FCC-he regions. The rectangle, in this figure,
represents the kinematics used for the top structure function into
the parameterization of the proton structure function. Indeed,
correlated bounds on the kinematic region at different values of
$Q^{2}$ are obtained from the parameterization of the proton
structure function, confronted with the HERA data. The kinematic
values are ($Q_{1}^{2}, Q_{2}^{2}, Q_{3}^{2})=(100, 1000,
3000)~\mathrm{GeV}^{2}$) for the LHeC and
FCC-he.\\
\begin{figure}[h]
\includegraphics[width=0.35\textwidth]{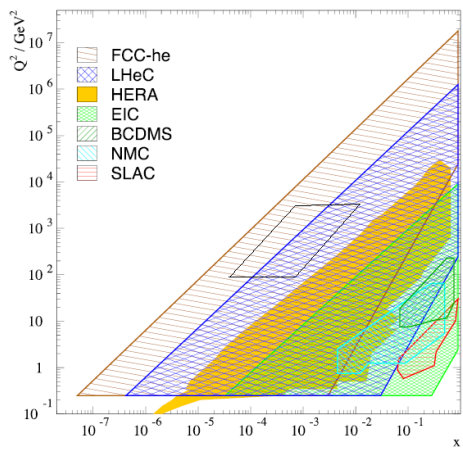}
\caption{The rectangle represents the coverage of the kinematic
plane in deep inelastic lepton -proton scattering by the LHeC and
FCC-he by extrapolating the HERA kinematic range to the top pair
production. The kinematics used are
$100~\mathrm{GeV}^{2}{\leq}Q^{2}{\leq}3000~\mathrm{GeV}^{2}$ and
$10^{-5}<x<10^{-2}$. Inelasticity is defined according to the CM
energies of ep colliders. }\label{Fig1}
\end{figure}
The top quark mass is $172{\pm}0.5~\mathrm{GeV}$, which has been
measured by ATLAS [21] and CMS [22] experiments. Here, $n_{f}=5$
is normally taken for $\mu^{2}<m^{2}_{t}$ at the top quark mass
threshold, where $\mu^{2}$ is the hard scale. Using $n_{f}=5$
active flavors in the running of $\alpha_{s}$ is important for the
precision phenomenology in the
 kinematics range
$100~\mathrm{GeV}^{2}{\leq}Q^{2}{\leq}3000~\mathrm{GeV}^{2}$.\\
The rest of our paper is organized as follows. In Sec. II, we
present our theoretical framework  of  the top structure function
into the parameterization of the proton structure function. In
Sec. III, we show the numerical results for the top distribution
and the implications to the measurement of the top cross section.
Finally, in Sec. IV , we present our conclusions.\\

\subsection{II. Theoretical framework}

The top quark structure functions in DIS in the further ep
colliders (i.e., LHeC and FCC-he) are obtained from the
measurements of the inclusive heavy quark cross sections, which
will be an important test of the QCD [1,2]. The reduced cross
section of the top quark is defined in terms of the top structure
functions by the following form:
\begin{eqnarray}
\sigma^{{t}\overline{{t}}}_{\mathrm{red}}(x,Q^{2})&=&
\frac{xQ^{4}}{2\pi
\alpha_{EM}^{2}[(1+(1-y)^{2}]}\frac{d^{2}\sigma^{{t}\overline{{t}}}}{dxdQ^{2}}\nonumber\\
&&=F_{2}^{t}(x,Q^{2})-f(y)F_{L}^{t}(x,Q^{2}),
\end{eqnarray}
where $f(y)=\frac{y^{2}}{1+(1-y)^{2}}$. The electron -proton
center of mass energies at the LHeC and FCC-he are proposed to be
$\sqrt{s}{\cong}1.3$ and $3.5~\mathrm{TeV}$, respectively. The
ratio $R^{t}(x,Q^{2})=F_{L}^{t}(x,Q^{2})/F_{2}^{t}(x,Q^{2})$ will
extend in future circular colliders (i.e., LHeC and FCC-he).
Indeed, these
new colliders are the ideal place to resolve this ratio [1].\\
The photon-proton cross section for the top pair production, in
the DIS, is related to the top structure function with the
following form
\begin{eqnarray}
\sigma_{2}^{t}(x,Q^{2})=4\pi^{2} \alpha_{EM}
F_{2}^{t}(x,Q^{2})/Q^{2}.
\end{eqnarray}
In the small $x$ range, where the gluon contribution is dominant,
the top structure functions in the collinear generalized DAS
approach are given by [17]
\begin{eqnarray}
F_{k}^{t}(x,Q^{2}){\simeq}~e^{2}_{t}\sum_{n=0}(\frac{\alpha_{s}}{4\pi})^{n+1}B^{(n)}_{k,g}(x,\xi)
{\otimes} xf_{g}(x,\mu^{2}),
\end{eqnarray}
where $k=2,L$ and $B_{k,g}$ are the collinear Wilson coefficient
functions in the high energy regime. Their compact forms are
defined in Ref.[17]. Here, $n$ denotes the order in running
coupling $\alpha_{s}$ and $\xi=\frac{m_{t}^{2}}{Q^{2}}$.\\
A NLO fit to the combined HERA data represents the power-like
behavior of the gluon distribution. The low $x$ asymptotic
behavior of $f_{g}(x,\mu^{2})$ is defined by the following form
[16,24]
\begin{eqnarray}
f_{g}(x,\mu^{2})|_{x{\rightarrow}0}{\rightarrow}\frac{1}{x^{1+\Delta}}.
\end{eqnarray}
We assume that this behavior holds for the gluon distribution
function, in the LHeC and FCC-he energy range, that creates the
top quark pair in the DIS processes [23]. According to the
Dokshitzer-Gribov-Lipatov-Altarelli-Parisi (DGLAP) evolution
equation, the singlet structure function at low $x$ is defined by
the gluon distribution as
\begin{eqnarray}
\frac{{\partial}F^{s}_{2}(x,Q^{2})}{{\partial}{\ln}Q^{2}}{\simeq}2n_{f}\sum_{n=0}(\frac{\alpha_{s}}{2\pi})^{n+1}
P^{(n)}_{qg}(x){\otimes}xf_{g}(x,\mu^{2}),
\end{eqnarray}
where $P_{qg}(x)$ is the quark-gluon splitting function. In Eqs.
(9) and (11), the symbol ${\otimes}$ is used for a shorthand
notation of the convolution formula, i.e.,
$f_{1}(x){\otimes}f_{2}(x){\equiv}\int_{x}^{1}\frac{dy}{y}f_{1}(y)f_{1}(\frac{x}{y})$.\\
After exploiting the low $x$ asymptotic behavior of the gluon
density (i.e., Eq.(10)), Eqs.(9) and (11) can be rewritten as
\begin{eqnarray}
F_{k}^{t}(x,Q^{2})&{\simeq}&M_{k,g}(x,\mu^{2},\Delta)xf_{g}(x,\mu^{2})
\end{eqnarray}
and
\begin{eqnarray}
\frac{{\partial}F_{2}(x,Q^{2})}{{\partial}{\ln}Q^{2}}&{\simeq}&N_{qg}(x,\mu^{2},\Delta)xf_{g}(x,\mu^{2}),
\end{eqnarray}
where
\begin{eqnarray}
N_{qg}(x,\mu^{2},\Delta)=\frac{5}{9}n_{f}\sum_{n=0}(\frac{\alpha_{s}}{2\pi})^{n+1}\int_{x}^{1}
P^{(n)}_{qg}(y)y^{\Delta-1}dy\nonumber\\
=\frac{5}{9}n_{f}\int_{x}^{1}
\bigg{[}\frac{\alpha_{s}}{2\pi}P^{(0)}_{qg}(y)+(\frac{\alpha_{s}}{2\pi})^{2}P^{(1)}_{qg}(y)\bigg{]}y^{\Delta-1}dy~~~~~~~
\end{eqnarray}
 and
\begin{eqnarray}
M_{k,g}(x,\mu^{2},\Delta)=e^{2}_{t}\sum_{n=0}(\frac{\alpha_{s}}{4\pi})^{n+1}\int_{x}^{x_{2}}
B^{(n)}_{k,g}(y,\xi)y^{\Delta-1}dy.\nonumber\\
=e^{2}_{t}\int_{x}^{x_{2}}
\bigg{[}\frac{\alpha_{s}}{4\pi}B^{(0)}_{k,g}(y,\xi)+(\frac{\alpha_{s}}{4\pi})^{2}B^{(1)}_{k,g}(y,\xi)\bigg{]}y^{\Delta-1}dy~~~
\end{eqnarray}
In the above equations (i.e., Eqs.(13-14) and (15)),
$F_{2}=\frac{5}{18}F_{2}^{S}(F_{2}^{S}=xf_{s})$ and
$x_{2}=1/(1+4\xi)$, respectively. Note that the superscript of the
coefficients on the right-hand side represents the order in
$\alpha_{s}$. The standard representation for QCD couplings in the
LO ($n=0$) and NLO ($n=1$) (within the
$\overline{\mathrm{MS}}$-scheme) approximations reads
\begin{eqnarray}
\alpha_{s}(t)=\frac{4\pi}{\beta_{0}t}~~~~~~~~~~~~~~~~~~~~~~~~~~(\mathrm{LO}),\nonumber\\
\alpha_{s}(t)=\frac{4\pi}{\beta_{0}t}\bigg{[}1-\frac{\beta_{1}
{\ln}{\ln}(t)}{\beta_{0}^{2}t}\bigg{]}~~~~~~(\mathrm{NLO}),\nonumber
\end{eqnarray}
with $\beta_{0}$ and $\beta_{1}$ as the first two coefficients of
the QCD $\beta$-function as
\begin{eqnarray}
\beta_{0}=\frac{1}{3}(11C_{A}-n_{f}),
~~~\beta_{1}=\frac{1}{3}(34C_{A}^{2}-2n_{f}(5C_{A}+3C_{F})),\nonumber
\end{eqnarray}
where $C_{F}=\frac{N_{c}^{2}-1}{2N_{c}}$ and $C_{A}=N_{c}$ are the
Casimir operators in the fundamental and adjoint representations
of the $\mathrm{SU(N_{c})}$ color group, and
$t=\ln\frac{Q^{2}}{\Lambda^{2}}$ where $\Lambda$ is the QCD
cut-off
parameter.\\
The top structure functions, with the gluonic dominant, i.e.,
Eq.(12), are obtained in the generalized DAS approach by the
following forms
\begin{eqnarray}
F_{k}^{t}(x,Q^{2})|_{\mathrm{g}}=\frac{M_{k,g}(x,\mu^{2},\Delta)}{N_{qg}(x,\mu^{2},\Delta)}
\frac{{\partial}F_{2}(x,Q^{2})}{{\partial}{\ln}Q^{2}},~k=2,L
\end{eqnarray}
where they are directly dependent on the derivative of the proton
structure function. In the LO approximation, Eq.(16) is
independent of the running coupling $\alpha_{s}$ and at the
high-order approximations, it depends on the running coupling.\\
By generalizing Eq.(11) to the singlet density, the singlet DGLAP
evolution equation is rewritten as
\begin{eqnarray}
\frac{{\partial}F^{s}_{2}(x,Q^{2})}{{\partial}{\ln}Q^{2}}&{=}&\sum_{n=0}(\frac{\alpha_{s}}{2\pi})^{n+1}\bigg{[}
P^{(n)}_{qq}(x){\otimes}xf_{s}(x,\mu^{2})\nonumber\\
&&+2n_{f}P^{(n)}_{qg}(x){\otimes}xf_{g}(x,\mu^{2})\bigg{]}.
\end{eqnarray}
The power -law behavior of the singlet distribution function is
introduced by the following form [16,24]
\begin{eqnarray}
f_{s}(x,\mu^{2})|_{x{\rightarrow}0}{\rightarrow}\frac{1}{x^{1+\Delta}},
\end{eqnarray}
where the singlet and gluon exponents in Eqs.(10) and (18) are
assumed to be equal at the NLO approximation\footnote{The Regge-
and non-Regge-like behavior assumptions for the gluon and singlet
distributions are explained in Ref.[25]}. Assuming the Regge-like
behavior for the gluon and singlet distributions (i.e., Eqs.(10)
and (18)), we obtain the following equation for the $Q^2$
derivative of the structure function $F_{2}$ as
\begin{eqnarray}
\frac{{\partial}F_{2}(x,Q^{2})}{{\partial}{\ln}Q^{2}}&{=}&T_{qq}(x,\mu^{2},\Delta)F_{2}(x,\mu^{2})\nonumber\\
&&+N_{qg}(x,\mu^{2},\Delta)xf_{g}(x,\mu^{2}),
\end{eqnarray}
where
\begin{eqnarray}T_{qq}(x,\mu^{2},\Delta)=\sum_{n=0}(\frac{\alpha_{s}}{2\pi})^{n+1}\int_{x}^{1}
P^{(n)}_{qq}(y)y^{\Delta-1}dy\nonumber\\
=\int_{x}^{1}
\bigg{[}\frac{\alpha_{s}}{2\pi}P^{(0)}_{qq}(y)+(\frac{\alpha_{s}}{2\pi})^{2}P^{(1)}_{qq}(y)\bigg{]}y^{\Delta-1}dy~~~~~~~
\end{eqnarray}
 Therefore, the top structure functions, with gluon and singlet distributions,
are obtained in the generalized DAS approach by the following
forms:
\begin{eqnarray}
F_{k}^{t}(x,Q^{2})|_{\mathrm{s}+\mathrm{g}}&{=}&\frac{M_{k,g}(x,\mu^{2},\Delta)}{N_{qg}(x,\mu^{2},\Delta)}
\bigg{[}\frac{{\partial}F_{2}(x,Q^{2})}{{\partial}{\ln}Q^{2}}\\
&&-T_{qq}(x,\mu^{2},\Delta)F_{2}(x,Q^{2})\bigg{]}.~k=2,L\nonumber
\end{eqnarray}
With the explicit form of the basic expression laid above (i.e.,
Eq.(21)), we can proceed to extract the top quark structure
functions $F_{2,L}^{t}$ from the parameterization of the proton
structure function at the LO and NLO approximations.\\
The parameterization of the proton structure function has been
found [18] at low values of Bjorken $x$ from fit to all of the
HERA DIS data. This function includes an asymptotic (high energy)
part that satisfies a saturated Froissart-bound behavior, with a
vector-dominance-like mass factor in the parameterization as
\begin{eqnarray}
F_{2}^{p}(x,Q^{2})=F_{2,\mathrm{v}}^{p}(x,Q^{2})
+F_{2,\mathrm{asymp}}^{p}(x,Q^{2})
\end{eqnarray}
where $F_{2,\mathrm{v}}^{p}$ is a valence term, and
$F_{2,\mathrm{asymp}}^{p}$ is an asymptotic term that is
 assumed to have  the Froissart-bounded form and reads
\begin{eqnarray}
F_{2,\mathrm{asymp}}^{p}(x,Q^{2})=D(Q^{2})(1-x)^{\nu}\sum_{m=0}^{2}A_{m}(Q^{2})L^{m},
\end{eqnarray}
where Eq.(23) is determined from the asymptotic part of the real
$\gamma^{*}p$ cross section by the following form:
\begin{eqnarray}
\sigma_{\mathrm{asymp}}^{p}&=&\lambda\frac{4\pi^{2}\alpha_{EM}}{M^{2}}
\Big {[}c_{0}+a_{0}\ln\Big {(}\frac{\nu}{m}\frac{2m^{2}}{\mu^{2}}
\Big {)}\nonumber\\
&&+b_{0}\ln^{2}\Big {(}\frac{\nu}{m}\frac{2m^{2}}{\mu^{2}}\Big
{)}\Big {]}.
\end{eqnarray}
The effective parameters in Eqs.(23) and (24) are defined in
Ref.[18]. This parameterization method, suggested  by the authors
in Ref.[18], provides reliable proton structure function
$F_{2}(x,Q^{2})$ with $x{\leq}0.1$ in a wide range of the momentum
transfer, $0.1~\mathrm{GeV}^{2}<
Q^{2}{\leq}5000~\mathrm{GeV}^{2}$.\\
A particular interest presenting the ratio of the top structure
functions, due to Eqs.(16) or (21), is defined as
\begin{eqnarray}
R^{t}(x,Q^{2})&=&\frac{F_{L}^{t}(x,Q^{2})}{F_{2}^{t}(x,Q^{2})}=\\
&&\frac{\sum_{n=0}(\frac{\alpha_{s}}{4\pi})^{n+1} \int_{x}^{x_{2}}
B^{(n)}_{L,g}(y,\xi)y^{\Delta-1}dy}
{\sum_{n=0}(\frac{\alpha_{s}}{4\pi})^{n+1} \int_{x}^{x_{2}}
B^{(n)}_{2,g}(y,\xi)y^{\Delta-1}dy}.\nonumber
\end{eqnarray}
This is comparable to those in the generalized DAS approach (see
[17]). Using the definition of the top structure function
$F_{2}^{t}$ in Eq.(21), the result for the $\gamma^{*}-p$ cross
section due to Eq.(8) is obtained by the following form:
\begin{eqnarray}
\sigma_{2}^{t}(x,Q^{2})&=&\frac{4\pi^{2}
\alpha_{em}}{Q^{2}}\frac{M_{2,g}(x,\mu^{2},\Delta)}{N_{qg}(x,\mu^{2},\Delta)}
\bigg{[}\frac{{\partial}F_{2}(x,Q^{2})}{{\partial}{\ln}Q^{2}}\nonumber\\
&&-T_{qq}(x,\mu^{2},\Delta){F_{2}(x,Q^{2})}\bigg{]}.
\end{eqnarray}
Therefore, it is straightforward to evaluate the $\gamma^{*}-p$
cross section using the parameterization
$F_{2}^{p}(x,Q^{2})$ and its derivatives.\\

\subsection{III. RESULTS}

We study the  top pair quark production in ep collisions at the
LHeC and FCC-he CM energies in the collinear generalized DAS
approach. Using the expression in Eq.(23) and extending the region
of data to lower $x$ values, according to Fig.1 in Ref.[18],
particular attention is paid to kinematics of low and ultra-low
values of the Bjorken variable $x$, $10^{-5}<x<10^{-2}$. The
validity of this extension could be checked in the future at the
proposed LHeC and FCC-he colliders. In further ep colliders, the
center-of-mass energy increases. Therefore, with the decrease in
the Bjorken variable $x$, it is possible to keep
$sx=\frac{Q^{2}}{y}$ constant.\\
The strong coupling constant $\alpha_{s}(M^{2}_{z})=0.118$ and the
running top quark mass $m_{t}=172.5~\mathrm{GeV}$ are used
throughout all the calculations. To estimate the scale
uncertainties of our calculations, the standard variations in
default renormalization and factorization scales, which were set
to be equal to $\mu_{R}^{2}={4m_{t}^{2}+Q^{2}}$ and
$\mu_{F}^{2}={Q^{2}}$, respectively, were introduced. The scale
variations are calculated by varying  the virtuality from
$Q^{2}=100~\mathrm{GeV}^{2}$ to $Q^{2}=3000~\mathrm{GeV}^{2}$.
Note that the  exponent $\Delta$ for the power-like behavior
$x^{-\Delta}$ of the collinear PDFs is considered to be
${\simeq}0.29$ or ${\simeq}0.50$ for the definition of the
uncertainties in our calculations. The value of the exponent
$\Delta{\simeq}0.29$ [26] agrees with the independent fit to the
DIS experimental data based on the hard
Pomeron conjecture leading to the exponent value $0.30{\pm}0.1$.\\
Our numerical results for the top structure function, the ratio
$R^{t}(x,Q^{2})$, and the top cross section are shown in Figs.2-4
at the renormalization scale $\mu_{R}^{2}$ for
$\Delta{\simeq}0.29$, respectively, in the LO and NLO
approximations. Results of calculations due to the CM energies and
comparison between the LHeC and FCC-he are presented
in these figures.\\
\begin{figure}[h]
\includegraphics[width=0.55\textwidth]{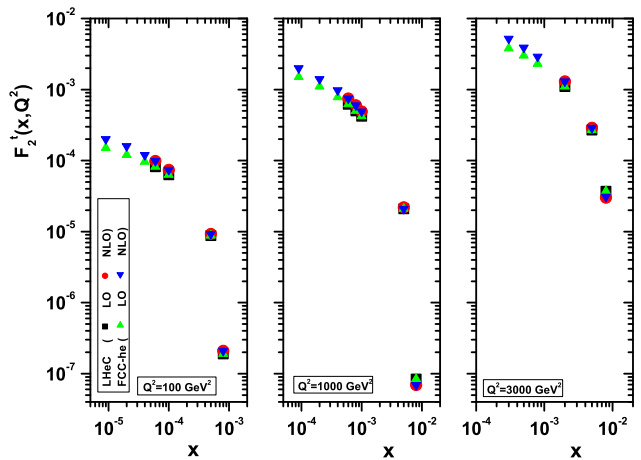}
\caption{ The $t\overline{t}$ structure function prediction in the
LHeC ($\sqrt{s}=1.3~\mathrm{TeV}$) and FCC-he
($\sqrt{s}=3.5~\mathrm{TeV}$) colliders at the LO and NLO
approximations for $Q^{2}=100,~1000$ and $3000~\mathrm{GeV}^2$ at
the renormalization scale $\mu_{R}^{2}={4m_{t}^{2}+Q^{2}}$ for
$\Delta{\simeq}0.29$, respectively. }\label{Fig2}
\end{figure}
In Fig.2, we present the $x$-dependence of the top structure
function $F_{2}^{t}(x,Q^{2})$ at $Q^{2}=100, 1000$, and
$3000~\mathrm{GeV}^{2}$ in the LO and NLO approximations. The top
structure functions increase as $x$ decreases. With respect to the
CM energies at the LHeC and FCC-he colliders, the results for the
FCC-he are larger than the LHeC at very low values of $x$.
Comparing the FCC-he and LHeC results, the FCC-he has a faster
growth rate. In other places, the results are comparable. We
observe that the top structure functions increase as $Q^{2}$
increases, and this is consistent with the pQCD.\\
In Fig.3, the $x$ dependence of the ratio $R^{t}$ is evaluated at
the LO and NLO approximations with $\mu_{R}^{2}=Q^{2}+4m^{2}_{t}$
for $\Delta{\simeq}0.29$.
\begin{figure}[h]
\includegraphics[width=0.55\textwidth]{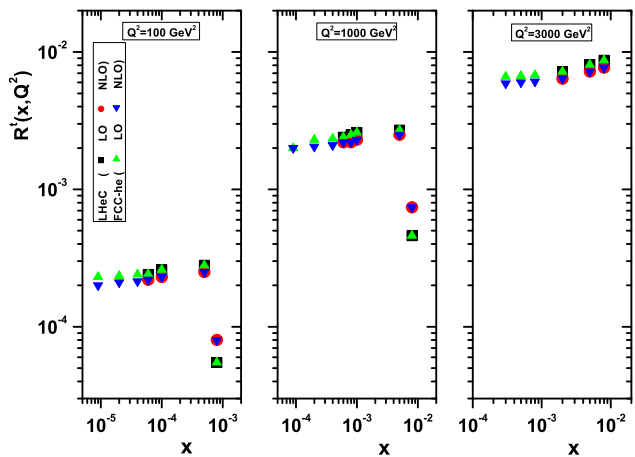}
\caption{ $R^{t}$ evaluated as functions of $x$ in the LHeC
($\sqrt{s}=1.3~\mathrm{TeV}$) and FCC-he
($\sqrt{s}=3.5~\mathrm{TeV}$) colliders at the LO and NLO
approximations with $\mu_{R}^{2}=Q^{2}+4m^{2}_{t}$ for
$Q^{2}=100,~1000$ and $3000~\mathrm{GeV}^{2}$.}\label{Fig3}
\end{figure}
These results lead to more or less flat (independent on $x$)
behavior of $R^{t}(x,Q^{2})$ with $R^{t}{\simeq}0.0002$ at
$Q^{2}=100~\mathrm{GeV}^{2}$ and $R^{t}{\simeq}0.006$ at
$Q^{2}=3000~\mathrm{GeV}^{2}$ values. There is a tendency for the
ratio $R^{t}(x,Q^{2})$ to be almost independent of $x$ in each bin
of $Q^{2}$ for very low $x$ and increasing with $Q^{2}$ increase
for $Q^{2}{\ll}m_{t}^{2}$, as it should be. The behavior of the
ratio $R^{t}(x,Q^{2})$ is in complete agreement with the usual
statement about the property of $k_{T}$-factorization for
$Q^{2}<5000~\mathrm{GeV}^{2}$, which grows
faster when $Q^{2}$ increases.\\
\begin{figure}[h]
\includegraphics[width=0.5\textwidth]{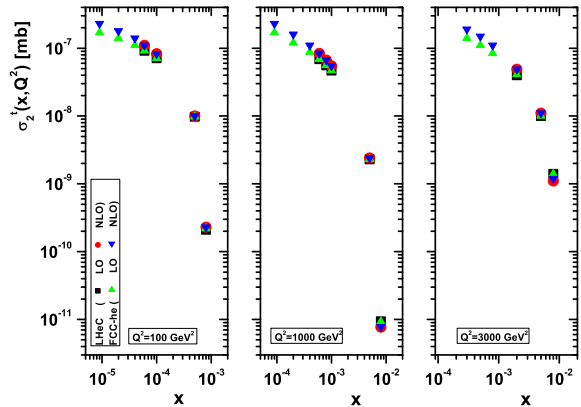}
\caption{ Behavior of the $t\overline{t}$ cross section as
function of $x$ at the LO and NLO approximations in the LHeC
($\sqrt{s}=1.3~\mathrm{TeV}$) and FCC-he
($\sqrt{s}=3.5~\mathrm{TeV}$) colliders  with
$\mu_{R}^{2}=Q^{2}+4m^{2}_{t}$ for $Q^{2}=100,~1000$ and
$3000~\mathrm{GeV}^{2}$. }\label{Fig4}
\end{figure}
\begin{figure}[h]
\includegraphics[width=0.5\textwidth]{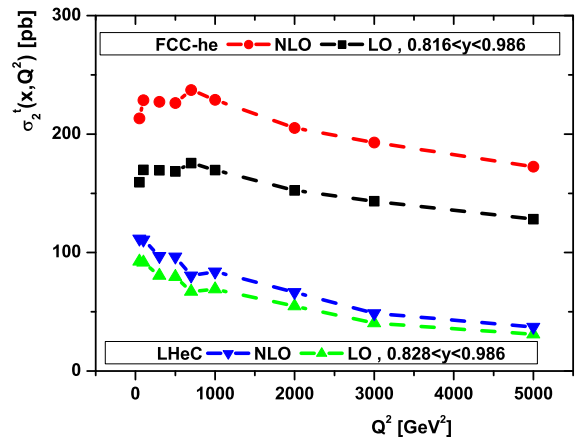}
\caption{ $\gamma^{*}p$ cross -sections $\sigma_{2}^{t}$ as
functions of $Q^{2}$ for fixed $\sqrt{s}$ for $0.8<y<1$ at the LO
and NLO approximations with $\mu_{R}^{2}=Q^{2}+4m^{2}_{t}$ for
$\Delta{\simeq}0.29$. }\label{Fig5}
\end{figure}
Photon -proton cross section for the top pair production is
related to the top structure function by Eq.(8). In figure 4, we
show results for $\sigma_{2}^{t}(x,Q^{2})$ as a function of $x$
for different $Q^{2}$ below $m_{t}^{2}$. Evolution of the
$t\overline{t}$ cross sections as a function of $x$ at the LO and
NLO approximations in the LHeC ($\sqrt{s}=1.3~\mathrm{TeV}$) and
FCC-he ($\sqrt{s}=3.5~\mathrm{TeV}$) colliders is shown in Fig.4.
These values increase as $x$ decreases. As a result, we predict
that at very low $x$ (i.e., at high inelasticity), the data will
increase at the FCC-he than the LHeC.\\
In figure 5, we plot $\gamma^{*}p$ cross sections for the LHeC and
FCC-he CM energies as functions of $Q^{2}$ at high inelasticity.
One can see that in both cases the cross -sections, for
inelasticity $0.8<y<1$, decrease as $Q^2$ increases. The $Q^2$
dependence of $\sigma_{2}^{t}$ in ep colliders is obtained at the
LO and NLO approximations with $\mu_{R}^{2}=Q^{2}+4m^{2}_{t}$ for
$\Delta{\simeq}0.29$ in this figure (i.e., Fig.5). It turns out
that these values at the LO and NLO approximations with the
renormalization scale numerically lead to
\begin{eqnarray}
30~\mathrm{pb}<\sigma^{t}_{2}({\mathrm{LHeC}})<130~\mathrm{pb},~0.828{\leq}y{\leq}0.986~~~~~~~~\nonumber\\
130~\mathrm{pb}<\sigma^{t}_{2}({\mathrm{FCC-he}})<250~\mathrm{pb},~0.816{\leq}y{\leq}0.986~~~
\end{eqnarray}
We observe that the difference between the LO and NLO results is
small and the extracted top cross sections are comparable for the
LHeC CM energy. This difference between the LO and NLO results is
large for the FCC-he CM energy. We observe that the results for
the top cross sections, for a fixed $Q^{2}$, at the FCC-he are
larger than the LHeC at the LO and NLO approximations. We observe
that the discrepancy between two considered future colliders tends
to be more clearly
pronounced at the NLO approximation.\\
To estimate the uncertainties of our calculations, the standard
variations in default scales (i.e., renormalization and
factorization ) and the behavior of the $\Delta$-exponents are
introduced. We present in Figs.6-8 the results for the top
properties at the NLO approximation with the FCC-he CM energy for
$Q^2=1000~\mathrm{GeV}^2$.
\begin{figure}[h]
\includegraphics[width=0.5\textwidth]{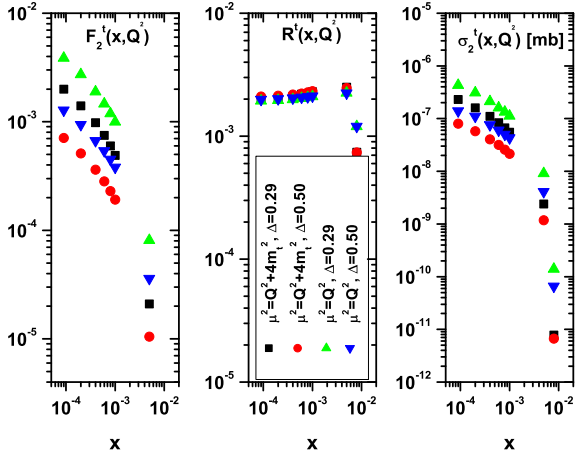}
\caption{ $x$ dependence of $F_{2}^{t}$, $R^{t}$ and
$\sigma_{2}^{t}$ at $Q^2=1000~\mathrm{GeV}^2$ for the FCC-he CM
energy at the NLO approximation for $0.8<y<1$. Plotted are
$\mu^2=Q^2+4m_{t}^2$ with $\Delta{\simeq}0.29$ (black -squares),
$\mu^2=Q^2+4m_{t}^2$ with $\Delta{\simeq}0.50$ (red -circles),
$\mu^2=Q^2$ with $\Delta{\simeq}0.29$ (green -up triangles) and
$\mu^2=Q^2$ with $\Delta{\simeq}0.50$ (blue -down
triangles).}\label{Fig6}
\end{figure}
The uncertainty range corresponds to the renormalization scale
$\mu^2=Q^2+4m_{t}^2$  with $\Delta{\simeq}0.29$ and $0.50$, and
the factorization scale $\mu^2=Q^2$ with $\Delta{\simeq}0.29$ and
${\simeq}0.50$, respectively. In Fig.6, the difference between
$F_{2}^{t}$ and $\sigma_{2}^{t}$ is clear when we compare these
results with the different scales and  exponents. One can see that
the results obtained with the factorization scale $\mu^2=Q^2$ with
$\Delta{\simeq}0.29$ are larger than others in a wide region of
$x$. Our predictions for the ratio of structure functions are
presented here, although they are fairly close.\\
\begin{figure}[h]
\includegraphics[width=0.5\textwidth]{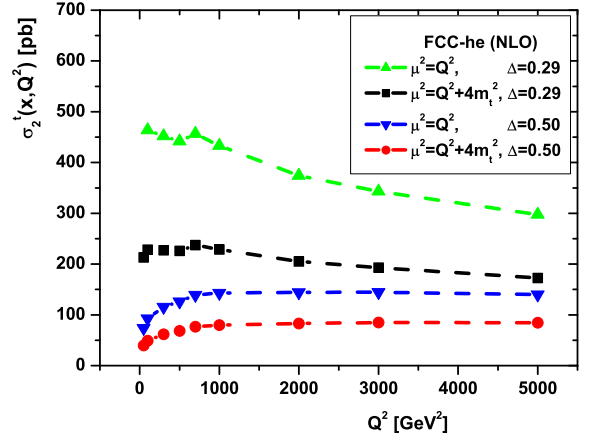}
\caption{ $Q^2$ dependence of $\sigma_{2}^{t}$ with the FCC-he CM
energy at the NLO approximation for $0.8<y<1$. Plotted are
$\mu^2=Q^2+4m_{t}^2$ with $\Delta{\simeq}0.29$ (black -squares)
and $\Delta{\simeq}0.50$ (red -circles), $\mu^2=Q^2$ with
$\Delta{\simeq}0.29$ (green -up triangles), and
$\Delta{\simeq}0.50$ (blue -down triangles). }\label{Fig7}
\end{figure}
Fig.7 shows the quantity $\sigma_{2}^{t}$ as a function of $Q^{2}$
for the FCC-he CM energy at the NLO approximation with respect to
the renormalization and factorization scales with
$\Delta{\simeq}0.29$ and ${\simeq}0.50$, respectively. One can see
that corrections due to the scales and exponents are sizable in a
wide range of $Q^2$. We see that, in this case, the factorization
scale with $\Delta{\simeq}0.29$ to $\sigma_{2}^{t}$ is strongly
different in comparison with the others.\\
\begin{figure}[h]
\includegraphics[width=0.5\textwidth]{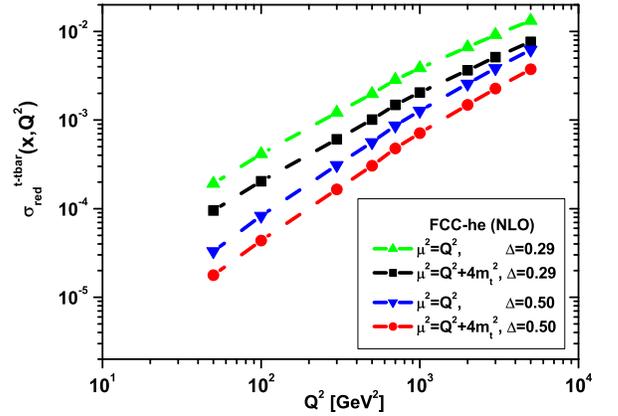}
\caption{ The reduced top cross sections
$\sigma_{\mathrm{red}}^{t\overline{t}}$ as a function of $Q^2$ at
the NLO approximation with the FCC-he CM energy for $0.8<y<1$
calculated at the scale $\mu^2=Q^2+4m_{t}^2$ with
$\Delta{\simeq}0.29$ (black -squares) and $\Delta{\simeq}0.50$
(red -circles), and at the scale $\mu^2=Q^2$ with
$\Delta{\simeq}0.29$ (green -up triangles) and
$\Delta{\simeq}0.50$ (blue -down triangles).}\label{Fig8}
\end{figure}
Our numerical results due to the theoretical uncertainties for
reduced cross section $\sigma_{\mathrm{red}}^{t\overline{t}}$ are
shown in Fig.8. In this figure (i.e., Fig.8), the $Q^2$ dependence
of the top reduced cross section is investigated at the NLO
approximation due to the FCC-he CM energy. The top reduced cross
sections increase as $Q^2$ increases. One can see again that the
results obtained with the scale $\mu^2=Q^2$ and
$\Delta{{\simeq}0.29}$ are larger than others in a wide region of
$x$ and $Q^{2}$. It is clear that the discrepancy between two
considered scale tends to be more clearly pronounced at small
$Q^{2}$. The uncertainties obtained for the top reduced cross
section extend into the range $10^{-5}<
\sigma_{r}^{t\overline{t}}\leq 10^{-2}$ for $50{\leq}Q^2{\leq}5000~\mathrm{GeV}^{2}$.\\

\subsection{IV. CONCLUSIONS}
In this paper, we present an overview of the top structure
function in future ep colliders that relies on the
Froissart-bounded parameterization of the proton structure
function in the collinear generalized DAS approach. We focus our
attention on the kinematic region of low $x$ in an interval of the
momentum transfer $Q^{2}<m^{2}_{t}$. The top structure functions,
the ratio of structure functions, and the top cross sections have
been considered within the LO and next-to-leading order (NLO)
approximations. The obtained explicit expressions for the selected
topics are entirely determined by the effective parameters of the
parameterization of the proton structure function. In principle,
due to the largess of the CM energies in the LHeC and FCC-he
colliders, the $x$-dominated region expands toward the lower
values while the $Q^{2}/y$ remains constant, and we can use the
HERA parameterization of the proton structure function that
describes fairly well the available experimental data on the
reduced cross sections.\\
The top cross sections extracted within the kinematic conditions
that will correspond to that will be available at the LHeC and
FCC-he colliders. It is found that, at relatively small
$Q^{2}<m^{2}_{t}$, both LO and NLO results reproduce the same and
uniform behavior well. We have performed an analysis of the
$x$-evolution of the extracted $F_{2}^{t}(x,Q^{2})$ and
$\sigma_{2}^{t}(x,Q^{2})$. We observe that the results increase as
$x$ decreases. The ratio of top structure functions, as a function
of $x$, is obtained for the LHeC and FCC-he CM energies at the LO
and NLO approximations. The behavior is in fairly good agreement
with the results [17] obtained in the framework of the
$k_{t}$-factorization method for the charm and bottom pair
production. Furthermore, the top structure functions can be used
to predict the top part of the neutrino-nucleon
cross -sections at ultra-high energy.\\
We obtained the uncertainties at high inelasticity on the top
properties from the collinear generalized DAS approach in the
kinematic range of the FCC-he collider at the NLO approximation.
We studied the dependence of the top reduced cross section on the
uncertainties due to the scales and exponents. It will be
interesting to compare these uncertainties with future results
from measurements of these top reduced cross sections in the
future. Carrying these results to improve the modeling of the top
quark properties is crucial toward probing optimally the
properties of this quark with the full data expected to be
interesting in the LHeC first, then in the FCC-he. Also, these
results may be important for future experiments at the
Electron-Ion Collider (EIC) and Electron-Ion Collider in
China (EiCC) [27] at low $x$.\\

\subsection{ACKNOWLEDGMENTS}
The author thanks  Razi University for financial support of this
project. The author also thanks the PLB referee for his/her
suggestions that helped improve the manuscript.

\section{References}

1. LHeC Collaboration and FCC-he Study Group, P. Agostini et al., J. Phys. G: Nucl. Part. Phys. {\bf48}, 110501(2021).\\
2. LHeC Study Group, J.L.Abelleira Fernandez et al.,
J. Phys. G: Nucl. Part. Phys. {\bf39}, 075001(2012).\\
3. M.Klein, arXiv [hep-ph]:1802.04317; M.Klein, Ann.Phys.{\bf528},
138 (2016); N.Armesto et al., Phys.Rev.D{\bf100}, 074022 (2019).\\
4. S.J.Brodsky, arXiv [hep-ph]: 1106.5820; H.Sun, arXiv [hep-ph]:
1710.06260; H.Denizli et al., Phys. Rev. D {\bf96}, 015024 (2017);
M.Gao and J.Gao, Phys. Rev. D {\bf104}, 053005 (2021).\\
5. Gerhard A. Schuler, Nucl.Phys.B {\bf299}, 21 (1988).\\
6. U. Baur and J.J. van der Bij, Nucl.Phys.B {\bf304}, 451
(1988).\\
7. M. Drees and K. Grassie, Z. Phys. C {\bf28}, 451 (1985); M.
Gl$\ddot{u}$ck and E. Reya, Phys. Lett.B {\bf83}, 98 (1979); L.M.
Jones and H.W. Wyld, Phys. Rev. D {\bf17}, 759 (1978).\\
8. N.Ya.Ivanov, Nucl.Phys.B {\bf814}, 142 (2009).\\
9. G.R.Boroun; Phys.Lett.B {\bf744}, 142 (2015); Phys.Lett.B
{\bf741}, 197 (2015); Chin.Phys.C {\bf41}, 013104 (2017);
 arXiv [hep-ph]:2109.09583; B.Rezaei and G.R.Boroun, EPL {\bf130}, 51002
 (2020).\\
10. Y.Kitadono, Phys.Lett.B {\bf702}, 135 (2011); Y.Yu et al.,
J.Phys.G {\bf45}, 125003 (2018); L.Han, Yan-Ju Zhang and Yao-Bei
Liu, Phys.Lett.B {\bf771}, 106 (2017).\\
11. R.D.Ball et al., Eur.Phys.J.C{\bf76}, 383(2016).\\
12. A.D.Martin, W.J.Stirling and R.G.Roberts, Phys.Rev.D{\bf50}, 6734(1994).\\
13. R.D.Ball and S.Forte, Phys.Lett.B{\bf335}, 77(1994); R.D.Ball and S.Forte, Phys.Lett.B{\bf336}, 77(1994).\\
14. A.Caldwell and M.Wing, Eur.Phys.J.C {\bf76}, 463(2016);
A.Caldwell et al., arXiv[hep-ph]:1812.08110(2018).\\
15. D.Britzger et al., Phys. Rev. D {\bf100}, 114007 (2019).\\
16. A.Yu.Illarionov, B.A.Kniehl and A.V.Kotikov, Phys. Lett. B
{\bf663}, 66 (2008); A.Yu.Illarionov, and A.V.Kotikov, Phys. Atom.
Nucl. {\bf75}, 1234 (2012); A.V.Kotikov et al., Lect.Notes Phys. {\bf647}, 386(2004); N.Ya. Ivanov, Nucl.Phys.B {\bf814}, 142 (2009).\\
17. A.V.Kotikov, A.V.Lipatov and P.Zhang, Phys.Rev.D {\bf104},
054042 (2021).\\
18. M. M. Block, L. Durand and P. Ha, Phys. Rev.D {\bf 89}, 094027 (2014).\\
19. L.P.Kaptari, A.V.Kotikov, N.Yu.Chernikova and Pengming Zhang,
Phys.Rev.D {\bf99}, 096019 (2019).\\
20. G.R.Boroun and B.Rezaei, Phys.Rev.D {\bf105}, 034002 (2022).\\
21. M.Aaboud et al. [ATLAS Collaboration], Eur.Phys.J.C {\bf79},
290 (2019).\\
22. V.Khachatryan et al. [CMS Collaboration], Phys. Rev. D
{\bf93},072004 (2016).\\
23. Olga B.Bylund, arXiv [hep-ex]:2103.14772;
 P.Ferreira da Silva, arXiv [hep-ph]:1605.05343.\\
24. J.R.Cudell and G.Soyez, Phys.Lett.B {\bf516}, 77 (2001);
A.Donnachie and P.V.Landshoff, Phys.Lett.B {\bf533}, 277 (2002);
Phys.Lett.B {\bf595}, 393 (2004).\\
25. A.V.Kotikov and G.Parente, Phys.Lett.B {\bf379}, 195 (1996).\\
26. D.Schildknecht, Phys.Rev.D {\bf104}, 014009 (2021);
G.R.Boroun, M.Kuroda and D.Schildknecht, arXiv
[hep-ph]:2206.05672.\\
27. A. Accardi et al., Eur. Phys. J. A {\bf52}, 268 (2016); D. P.
Anderle et al., Frontiers of Physics {\bf16}, 64701 (2021).\\

\end{document}